\documentclass[conference]{IEEEtran}
\IEEEoverridecommandlockouts

\usepackage{cite}
\usepackage{amsmath,amssymb,amsfonts}
\usepackage{algorithmic}
\usepackage{graphicx}
\usepackage{textcomp}
\usepackage{xcolor}
\usepackage{cite}
\usepackage[linesnumbered,ruled,vlined]{algorithm2e}
\usepackage{enumerate}
\usepackage{subcaption}
\def\BibTeX{{\rm B\kern-.05em{\sc i\kern-.025em b}\kern-.08em
    T\kern-.1667em\lower.7ex\hbox{E}\kern-.125emX}}

\usepackage[top=0.76in, bottom=1.02in, left=0.65in, right=0.65in]{geometry}
\begin{document}
\bstctlcite{IEEEexample:BSTcontrol}

\setlength{\belowcaptionskip}{2pt}
\setlength{\abovecaptionskip}{2pt}
\setlength{\textfloatsep}{6pt}
\setlength{\floatsep}{6pt}

\title{Joint Fiber and Free Space Optical Infrastructure Planning for Hybrid Integrated Access and Backhaul Networks
{\footnotesize \textsuperscript{}}
\\\textit{(Invited Paper)}\\
}
%alternative author block below

\author{
\IEEEauthorblockN{Charitha Madapatha, Piotr Lechowicz, Carlos Natalino, Paolo Monti, Tommy Svensson}
\IEEEauthorblockA{
Department of Electrical Engineering, Chalmers University of Technology, Gothenburg, Sweden. \\
\{charitha, piotr.lechowicz, carlos.natalino, mpaolo, tommy.svensson\}@chalmers.se}
}

\maketitle

\begin{abstract}
Integrated access and
backhaul (IAB) is one of the promising techniques for 5G networks and beyond (6G), in which the same node/hardware is used to provide both backhaul and cellular services in a multi-hop architecture. Due to the sensitivity of the backhaul links with high rate/reliability demands, proper network planning is needed to ensure the IAB network performs with the desired performance levels. In this paper, we study the effect of infrastructure planning and optimization on the coverage of IAB networks. We concentrate on the cases where the fiber connectivity to the nodes is constrained due to cost. Thereby, we study the performance gains and energy efficiency in the presence of free-space optical (FSO) communication links. Our results indicate hybrid fiber/FSO deployments offer substantial cost savings compared to fully fibered networks, suggesting a beneficial trade-off for strategic link deployment while improving the service coverage probability. As we show, with proper network planning, the service coverage, energy efficiency, and cost efficiency can be improved.
%TODO: complete
\end{abstract}


\section{Introduction}
The rapid growth in mobile data traffic, fueled by the rise of smart devices, immersive applications, and the Internet of Things (IoTs), has significantly increased the demand for high-speed, low-latency, and ultra-reliable communication networks. To meet this growing need, network densification has emerged as a key strategy in the evolution of 5G and the development toward future 6G \cite{2021:Alasabah:Wireless_Communications_Networks}. This approach involves deploying many low-power, small base stations (SBSs), access points (APs) or reconfigurable intelligent surfaces (RISs) within a geographical area. The primary objectives are enhancing spatial frequency reuse, increasing capacity, and reducing end-to-end latency by bringing the network infrastructure closer to end users \cite{10757793}. Realizing the full benefits of network densification introduces challenges concerning backhauling, i.e., the process of connecting distributed access points to the core network. Traditional backhaul solutions, such as fiber optics and microwave links, each come with their own advantages and disadvantages \cite{2016:Jaber:5G_Backhaul_Challenges}.

Fiber optics offer high throughput and reliability but have high deployment costs and extended installation times, posing challenges for rapid expansion in dense urban areas or hard-to-reach rural locations \cite{2012:AhmadAnas:Hybrid_FTTX_FSO}. On the other hand, microwave links offer faster deployment and greater flexibility in network topology. However, they are constrained by the availability of licensed spectrum and can be affected by interference and weather-related fading, particularly in the millimeter-wave (mmWave) bands, which may impair performance \cite{2020:Hassan:Channel_Estimation_Techniques}.
To tackle these challenges, Integrated Access and Backhaul (IAB) has emerged as a crucial solution for cost-effective and scalable network expansion. Standardized in 3GPP Release 16 and later, IAB uses the same radio access technology and spectrum for both user access and backhaul transmission, significantly reducing the need for wired infrastructure. IAB nodes operate as relay points, creating multi-hop wireless mesh topologies that extend coverage and capacity in a self-organizing and adaptive way. However, despite these advantages, IAB systems are still susceptible to the limitations of wireless backhaul links, especially in situations involving line-of-sight (LoS) blockages, weather conditions (e.g., rain fade), and complexities introduced by dynamic urban environments \cite{ madapatha2020integrated}.

In response to these vulnerabilities, researchers and industry stakeholders have focused on hybrid backhaul architectures, more specifically those where IAB is integrated with Free-Space Optical (FSO) communication and fiber. FSO systems use tightly focused light beams, typically in the infrared or visible spectrum, to transmit high-speed data wirelessly without needing physical cables. These systems offer several significant advantages, including multi-Gbps throughput, immunity to electromagnetic interference, enhanced security due to their narrow beamwidths, and operation within license-free spectrum bands. Additionally, employing FSO links can alleviate congestion in the radio frequency (RF) spectrum and serve as a complementary backhaul path for mmWave IAB systems, providing a high-capacity alternative, especially where a fiber deployment is too costly.

FSO links are highly vulnerable to atmospheric conditions (e.g., fog, heavy rain, dust), which can lead to significant signal loss and outages. To address these challenges, advanced adaptive link management strategies have been proposed. These include dynamic on-off switching, load balancing, and dual-mode transceivers that can switch between FSO and RF backhaul based on real-time link quality metrics and environmental sensing. These strategies enhance the backhaul network's resilience and reliability while allowing for context-aware resource allocation and quality of service (QoS) provisioning \cite{2023:Mohsan:Hybrid_FSO_RF}.

Although there is growing interest in IAB and FSO technologies, much of the existing literature treats them separately. Studies focused on IAB usually examine aspects such as optimal node placement, routing algorithms, and scheduling, often under idealized channel conditions and without considering the integration of optical links \cite{9864179,10119093}. On the other hand, research related to FSO emphasizes channel modelling, beam alignment, and techniques for atmospheric compensation, but not many address how these systems can be incorporated into broader radio access network (RAN) architectures. This fragmented approach creates a significant gap in developing comprehensive hybrid fiber/FSO backhaul solutions.
To address this issue, we propose a unified and intelligent backhaul framework, one that seamlessly integrates IAB, FSO, and fiber optics based on real-time conditions, network demands, and service requirements. This framework would utilize the strengths of each technology, i.e., to use the widespread applicability and scalability of IAB, the ultra-high capacity of FSO, and the reliability of fiber. By implementing such solutions, next-generation wireless networks can achieve enhanced performance, adaptability, sustainability, and cost-effectiveness.

In this paper, we study the effect of network planning on the service coverage probability of IAB networks assisted with fiber/FSOs.
We present fiber/FSO deployment algorithms that jointly optimize the inter-node fiber distance and cost. 
Moreover, we study the effect of different parameters on the network performance focusing on the service coverage probability and energy efficiency. As we show, the coverage and cost efficiency of IAB networks can be considerably improved via proper network planning.

% \subsection{Architecture}
% In an IAB network, multi-hop communication is enabled through a hierarchical chain of IAB nodes connected to a central IAB donor. The IAB donor, which interfaces with the core network via a high-capacity link (e.g., fiber), hosts the Central Unit (CU) that manages downstream IAB nodes. Each IAB node comprises two functional entities: the Distributed Unit (DU) and the Mobile Termination (MT). The DU provides gNB-like functionalities and is responsible for serving user equipments (UEs) or the MT of downstream (child) IAB nodes. Conversely, the MT component establishes a wireless backhaul link with the DU of an upstream (parent) IAB node, thus forming a multi-hop path toward the IAB donor.

% While the DU typically aligns with standard gNB operations, the MT may exhibit IAB-specific capabilities to support backhaul functions. However, functionally, the MT operates similarly to a UE from the perspective of its parent DU, enabling seamless integration into the RAN stack. This architecture supports flexible and scalable deployments, especially when direct fiber connectivity to all small cells is infeasible.

\begin{figure}
\centerline{\includegraphics[width=2.75in]{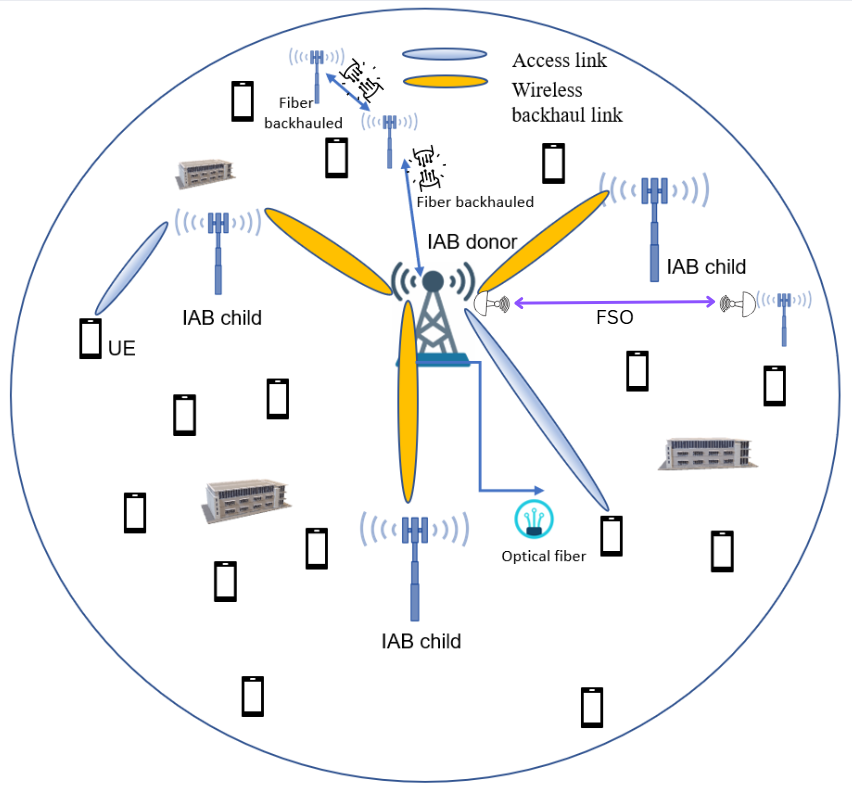}}
\caption{\textcolor{black}{Schematic of the hybrid IAB system model.}}
\label{systemmodel}
\end{figure}

\section{System Model}

We consider the downlink of a two-hop IAB network, where an MBS, i.e., IAB donor and its associated SBS, i.e., IAB child nodes serve a set of UEs as illustrated in Fig.~\ref{systemmodel}~\cite{singh2015tractable,saha2019millimeter,madapatha2020integrated,adare2021uplink}. Particularly, finite homogeneous Poisson point processes (FHPPPs) are used to model the spatial distributions of the MBSs, SBSs, and UEs. An in-band mmWave communication framework is assumed, where both access and backhaul links share the same frequency band. While this setup increases coordination complexity, it offers flexible spectrum utilization without requiring dedicated frequency channels. We adopt the LoS probability model for urban macro (UMa) scenario as detailed in 3GPP TR
38.901,~\cite{3gpp2017channelmodel}, for blockage, and the 5GCM UMa close-in channel model~\cite{ref3,madapatha2020integrated} to characterize mmWave propagation. The received power at node $r$ from transmitter $t$ is modeled as:
\begin{equation}
    P_{\mathrm{rx}} = P_{\mathrm{tx}} h_{t,r} G_{t,r} L_{t,r}(\|x_t - x_r\|)^{-1},
    \label{eq:rx_power}
\end{equation}
where $P_{\mathrm{tx}}$ is the transmit power, $h_{t,r}$ denotes small-scale fading modelled as a normalized Rayleigh
random variable, $G_{t,r}$ is the combined antenna gain, and $L_{t,r}(\cdot)$ represents path loss, which depends on LoS or Non-LoS (NLoS) conditions determined by the blockage model.

The antenna gain $G_{t,r}$ follows a simplified sector antenna pattern:
\begin{equation}
G_{t,r}(\theta) = 
\begin{cases}
G_\mathrm{main}, & |\theta| \leq \frac{\theta_\mathrm{HP}}{2} \\
G_\mathrm{side}, & \text{otherwise}
\end{cases},
\end{equation}
where $\theta_\mathrm{HP}$ is the half-power beamwidth, and $G_\mathrm{main}$, $G_\mathrm{side}$ denote the main-lobe and side-lobe gains, respectively \cite{ref1}.

Each UE $u$ associates with either the IAB donor or one of the child IAB nodes based on received power. The interference at UE $u$ is defined as:
\begin{equation}
I_u = \sum_{i \in \Phi_u \setminus \{w_u\}} P_i h_{i,u} G_{i,u} L_{i,u}(\|x_i - x_u\|)^{-1},
\label{eq:interference_ue}
\end{equation}
where $w_u$ is the serving node of UE $u$, and $\Phi_u$ is the set of all potential interferers.

Similarly, the aggregate interference experienced by a child IAB node $c$ on its backhaul link is:
\begin{equation}
I_c = \sum_{j \in \Phi_c \setminus \{w_c\}} P_j h_{j,c} G_{j,c} L_{j,c}(\|x_j - x_c\|)^{-1},
\label{eq:interference_iab}
\end{equation}
where $w_c$ denotes the IAB donor serving the IAB child node $c$. The available mmWave bandwidth $W$ is partitioned between access and backhaul links as:
\begin{equation}
\begin{aligned}
W_{\mathrm{bh}} &= \beta W, \\
W_{\mathrm{ac}} &= (1 - \beta) W,
\end{aligned}
\label{eq:bandwidth_split}
\end{equation}
where $\beta \in [0,1]$ denotes the backhaul bandwidth allocation factor.

% \addtolength{\topskip}{2mm}
Let $\mathcal{D}$, $\mathcal{C}$, and $\mathcal{U}$ denote the sets of IAB donors, child IAB nodes, and UEs, respectively. Each UE $u \in \mathcal{U}$ may be served either directly by the IAB donor or through a child IAB node. Accordingly, the downlink rate $R_u$ for UE $u$ is given by:
\begin{equation}
R_u =
\begin{cases}
\displaystyle\frac{W_{\mathrm{ac}}}{N_d} \log_2(1 + \mathrm{SINR}_u), & w_u \in \mathcal{D},\ \\[6pt]
\min\Bigg\{
\begin{aligned}
&\displaystyle\frac{W_{\mathrm{ac}}}{N_{c,u}} \log_2(1 + \mathrm{SINR}_u)\\
&\displaystyle\frac{W_{\mathrm{bh}}}{N_{c}} \log_2(1 + \mathrm{SINR}_c)
\end{aligned}
\Bigg\}, & w_u \in \mathcal{C}
\end{cases},
\label{eq:rate_expression}
\end{equation}
where $N_d$ is the number of UEs directly served by the donor, $N_{c,u}$ is the number of UEs served by child IAB node $c$ (to which $u$ is associated), and $N_c$ is the total number of child IAB nodes. In addition, for UEs served by SBSs with non-IAB backhaul (e.g., fiber or FSO-connected), the downlink rate is expressed as:

\begin{equation}
R_u = \frac{W}{N_u} \log_2(1 + \mathrm{SINR}_u), \quad \text{if } w_u \in \mathcal{S},
\label{eq:fibsbs}
\end{equation}
where $N_u$ denotes the total number of UEs connected to the \textcolor{black}{non-IAB backhauled} SBS of which the considered UE is associated. 

Our design objective is to optimize node deployment and bandwidth allocation to maximize service coverage probability, given by \( \Pr(R_u \geq \eta) \) and  defined as the fraction of UEs whose instantaneous rate exceeds a target threshold $\eta$.

% \begin{table*}[t]
% \centering
% \caption{The Definition of the Parameters.}
% \label{table}
% \setlength{\tabcolsep}{3pt}
% \begin{tabular}{|p{40pt}|p{160pt}|p{40pt}|p{160pt}|}
% \hline
% Parameter& 
% Definition&
% Parameter&
% Definition\\
% \hline

%  $\lambda_{\text{bl}}$& Blocking density&$\lambda_{\text{T}}$& Tree density \\
%  $\theta$& Orientation of the blocking wall&$\alpha$&Angle between transmitter and receiver\\
%  $P_{\text{t}}$&Transmission power&$P_{\text{r}}$&Received power\\
%  $h$&Fading coeficient&$G$&Antenna gain\\

%  $x$&Location of the node&$r$&Propagation distance  between  the  nodes\\
 
% \hline
% \end{tabular}
% \label{tab1x}
% \end{table*}

\begin{algorithm}[tbph]
\caption{Plan connected fiber-backhaul topology}
\label{alg:connected-topology}
% \SetKwFunction{FConnectedGraph}{connected\_graph}
% \SetKwProg{Fn}{Fn}{:}{}
%
\DontPrintSemicolon
\KwIn{$\Gamma = (V, E)$: Initial disconnected graph (e.g., only MBS ring)}
\KwOut{
$\Gamma$: Connected graph with all nodes reachable}
$V_0 \gets \{v \in V : \deg(v) = 0\}$ \tcp*{Nodes with no connections}
\While{$V_0 \neq \emptyset$}{
    Select random node $v \in V_0$ \;
    $u \gets \arg\min_{w \in V \setminus V_0} dist(v, w)$ \tcp*{Closest connected node}
    Add edge $(v, u)$ to $\Gamma$\;    
    $V_0 \gets V_0 \setminus \{v\}$\;
}
\KwRet $\Gamma$ \;
\end{algorithm}

\begin{algorithm}[tbph]
\caption{Fiber-backhaul connection placement FBCP($\alpha$)
% Fber-backhaul connection placement with minimum inter-node distance between fiber connected nodes
}
\label{alg:nodes-selection}
% \textcolor{blue}{Piotr -- Algorithms minimum fiber node}
\DontPrintSemicolon
\KwIn{
$\Gamma = (V, E)$: network topology with MBS ring; $V_m$: MBSs; $V_{IAB} \subset V_s$ SBSs with IAB backhaul; $V_{Fib} \subset V$ SBSs with fiber backhaul; $N$: number of SBSs to connect with fiber;
$\alpha \in [0, 1]$: Weight parameter for separation vs. cost}
\KwOut{Updated graph $\Gamma$ with $N$ additional fiber-connected SBSs}

$\Gamma_{design} \gets$ Connected reference graph from Algorithm~\ref{alg:connected-topology} \;

\For{$i \gets 1$ \KwTo $N$}{
    \ForEach{$v \in V_{IAB}$}{
        $p^* \gets$ null \tcp*{Best path to an MBS}
        \ForEach{$v_m \in V_m$}{
            $p_{v, v_m} \gets$ Shortest path from $v$ to $v_m$ in $\Gamma_{design}$ \;
            
            \If{$p^* = \text{null}$ \textbf{or} $l(p_{v, v_m}) < l(p^*)$}{
                $p^* \gets p_{v, v_m}$ \;
            }
            
        }
        $E^\prime \gets$ Edges in $p^*$ not already in $\Gamma$ \;

        $\Gamma^\prime[v] \gets$ Updated graph  $\Gamma$ with edges $E^\prime$ \;
        
        $cost[v] \gets$ $\beta_{dig} \cdot l(E^\prime) + \beta_{fiber} \cdot l(p^*) + 2 \cdot \beta_{trx}$, cost of moving from $\Gamma$ to $\Gamma^\prime$\;
           
        $sep[v] \gets$
        $\sum_{u \in V_{Fib} \cup V_{m}} \frac{1}{dist(v, u)}$
        \;

        % \If{$c < c^*$}{
        %     $c^* \gets c$\;
        %     $v^* \gets v$ \;
        %     $\Gamma^* \gets \Gamma^\prime$ \;
        % }
    }
    Normalize $cost[v]$ and $sep[v]$ for all $v \in V_{IAB}$ \;

    \ForEach{$v \in V_c$}{
    $weight[v] \gets \alpha \cdot sep[v] + (1 - \alpha) \cdot cost[v]$\;
    }
    $v^* \gets \arg\min_{v \in V_{IAB}} weight[v]$ \;

    $\Gamma \gets \Gamma^\prime[v^*]$\;
    $ V_{IAB} \gets V_{IAB} \setminus \{ v^* \} $ \;
    $ V_{Fib} \gets V_{Fib} \cup \{ v^* \} $ \;
}
\KwRet $ \Gamma, V_{IAB}, V_{Fib}$\;
% \textit{Return the set of the node locations in Step V as the optimal node location set.}
\end{algorithm}

% \begin{algorithm}[tbph]
% \caption{Fiber-backhaul connection placement minimizing cost
% % Fber-backhaul connection placement with minimum inter-node distance between fiber connected nodes
% }
% \label{alg:fiber-connected-cost}
% % \textcolor{blue}{Piotr -- Algorithms minimum fiber node}
% \DontPrintSemicolon
% \KwIn{Graph $\Gamma = (V, E)$ -  network topology with ring connecting MBSs $V_m$, $V_c \in V$ fiber-disconnected IAB children nodes, $V_d \in V$ fiber-connected IAB donor nodes, $N$ - number of nodes to fiber connect}
% \KwOut{Graph $\Gamma$ connecting given number of nodes to form IAB donors}
% $\Gamma_{design} \gets$ connected graph from $\Gamma$ using Algorithm 1\;
% \For{$i = 1$ to $N$}{
%     $v^*, c^*, \Gamma^* \gets$ null, null, null \;
%     \For{$v \in V_c$}{
%         $\Gamma^\prime \gets$ create topology from $\Gamma$ by adding edges included in shortest path from $v$ to any node in $V_m$ in $\Gamma_{design}$ if they are not already in $\Gamma$\;
%         $c \gets$ cost of $\Gamma^\prime$ \;  
%         \If{$c < c^*$}{
%             $c^* \gets c$\;
%             $v^* \gets v$ \;
%             $\Gamma^* \gets \Gamma^\prime$ \;
%         }
%     }
%     $\Gamma \gets \Gamma^*$\;
%     $ V_c \gets V_c \setminus \{ v^* \} $ \;
%     $ V_d \gets V_d \cup \{ v^* \} $ \;
%     \KwRet $ \Gamma, V_c, V_d$\;
% }
% % \textit{Return the set of the node locations in Step V as the optimal node location set.}
% \end{algorithm}

\subsection{Fiber Backhaul Connection Placement Algorithm}

We model the base station fiber-backhaul planning as a graph $\Gamma = (V, E)$, where $V$ represents the set of base stations and $E$ the set of fiber links interconnecting them. Base stations are classified as main base stations (MBSs) $V_m$ or small base stations (SBSs) $V_s$. SBSs may be connected via either IAB $V_{IAB} \subset V_s$ or fiber $V_{Fib} \subset V_s$. MBSs $V_m$ are interconnected by fiber links $E_m \subset E$ that form a ring topology and serve as IAB donor nodes. Let $l(P)$ denote the total length of fibers along path $P$, and let $dist(v, u)$ denote the Euclidean distance between base stations $v$ and $u$.
% Each IAB connected with fiber can become a donor node for other child IAB, each disconnected node serves as a child IAB. The network contains $V_m$ macro base stations (MBSs) with wider range and processing capabilities. 

Although the long-term goal is to connect all nodes via fiber, in practice, this deployment occurs incrementally to amortize the cost over time. The eventual fiber-connected network is designed using the Algorithm~\ref{alg:connected-topology}, which incrementally expands an initially connected subgraph. Disconnected nodes are processed in random order and connected to their nearest already connected neighbor. In our case, the initial subgraph is the ring topology formed by the MBSs $V_m$. The aim is to minimize the deployment cost and maximize the average SNR, which both are scaling with respect to distance \cite{2021:Matzner:Making_Intelligent_Topology}.

We assume the primary costs in optical network deployment arise from trenching, optical fiber, and transceivers. The MBSs $V_m$ are interconnected in a ring topology, while each fiber-connected SBS $v \in V_{Fib}$ is linked to its nearest MBS via a dedicated point-to-point fiber. The full deployment follows the structure produced by Algorithm~\ref{alg:connected-topology}. We further assume that multiple fiber links can share a trench. All cost metrics are normalized relative to the cost of a 10G transceiver ($\beta_{trx}=1$), with trenching cost set at $\beta_{dig}=2.4$ units per meter and fiber cost at $\beta_{fiber}=0.006$ units per meter \cite{2020:Udalcovs:TotalCostOwnership}. 
We express the potential cost of an FSO link within the range $\beta_{FSO}=\lbrack1, 100\rbrack$. 

While the long-term infrastructure plan assumes full fiber connectivity, mid-term decisions about which SBSs should be fiber-connected remain unresolved. To address this, we propose 
%a strategy that focuses jointly on minimizing deployment cost and maximizing spatial separation among fiber-connected SBSs to improve coverage
% two strategies: one that prioritizes minimizing deployment cost, and another that aims to maximize spatial separation among fiber-connected SBSs to improve coverage.
Fiber-Backhaul Connection Placement algorithm (FBCP, Algorithm~\ref{alg:nodes-selection})  connects $N$ SBSs with fiber links, guided by a weighting parameter $\alpha$. This parameter balances minimizing deployment cost ($\alpha$=0) and maximizing the spatial separation among fiber-connected SBSs ($\alpha$=1) to improve coverage. The algorithm starts with the initial ring topology $\Gamma$ composed of the MBSs $V_m$ and incrementally connects SBSs until $N$ nodes are fiber-connected.

At each step, the algorithm evaluates all currently unconnected SBSs $V_{IAB}$ (line 3). For each node $v \in V_{IAB}$, it computes the shortest path to the closest IAB donor in $V_m$, using the design graph $\Gamma_{design}$ (lines 4–8). Let $E^\prime$ denote the set of edges required to connect node $v$ to the existing graph $\Gamma$. The cost of connecting $v$ is computed in line 11. The inverse separation is calculated as the sum of inverse Euclidean distances from node $v$ to all currently fiber-connected nodes (line 12). After normalizing both cost and separation metrics, the node minimizing the weighted objective is selected (lines 13–16). This node is then connected via fiber according to $\Gamma_{design}$, and the process repeats until $N$ SBSs are connected.

\section{Simulation Results and Discussion}
\label{res_sec}

In this section, we evaluate the effect of fiber/FSO deployment strategies on the cost, service coverage probability, and energy efficiency of the IAB networks. 
Here, the nodes are connected according to FBCP (Algorithm~\ref{alg:nodes-selection}) with separation-cost weight parameter $\alpha$, where $\alpha$=1 corresponds to scenario focused on increasing nodes separation solely 
%(and $\alpha$=0 on minimizing cost).
and \textit{cost sort}, i.e., minimizing cost with $\alpha$=0). We consider an IAB network operating at 28 GHz, with $W$
= 1 GHz, $G_\mathrm{main}$ = 24 dBm, $G_\mathrm{side}$ = -2 dBm and $\eta$ = 100 Mbps. The network consists 5 MBSs, i.e., IAB donors, 80 SBSs, i.e., IAB child nodes and 1000 UEs, for the evaluations, unless otherwise stated.

\subsection{Cost Behavior of Fiber-Connected SBSs}

\begin{figure}
\centering
\includegraphics[width=\linewidth, clip=false]{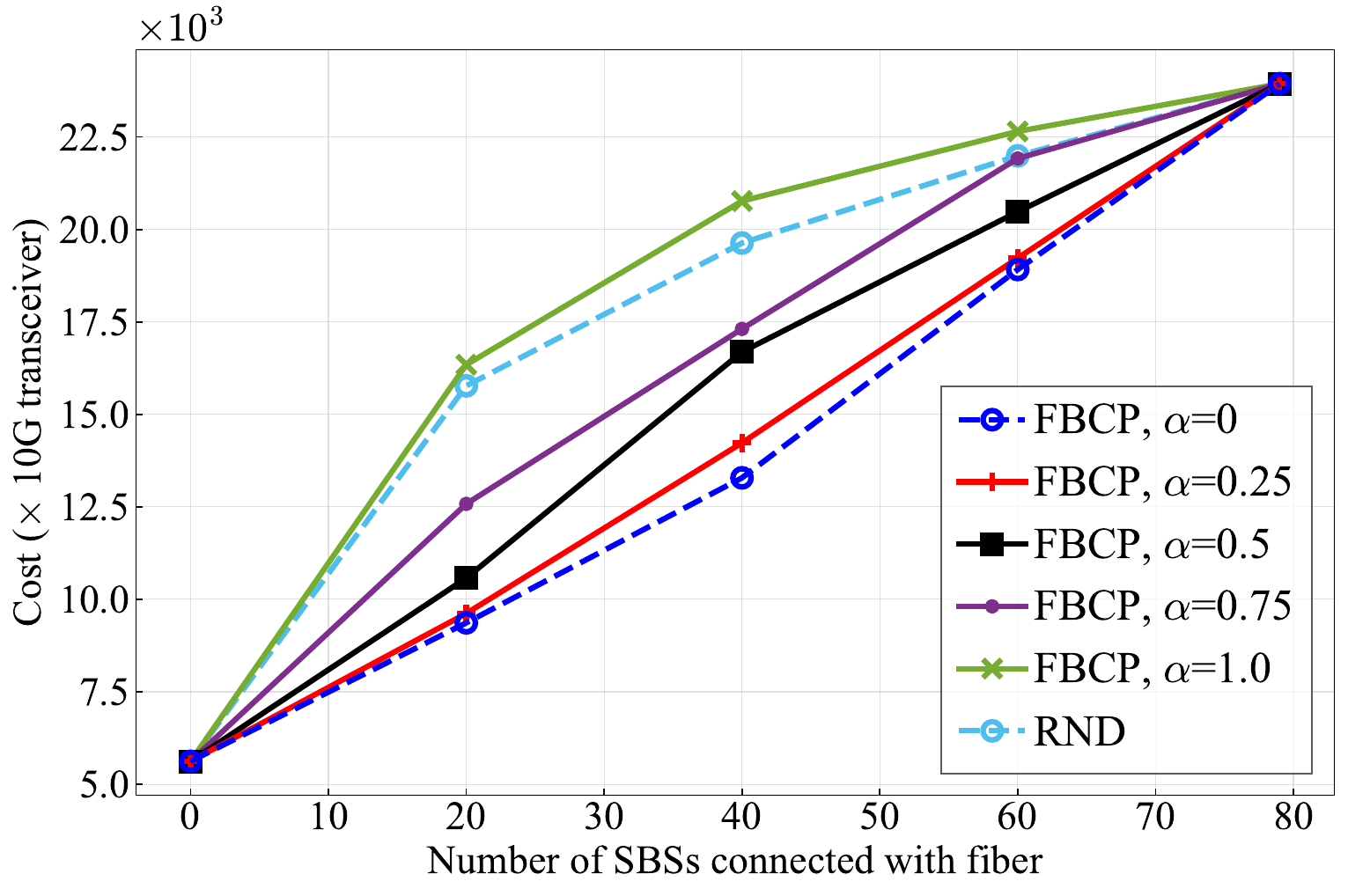}
\caption{\textcolor{black}{Cost as a function of the number of SBSs connected with fiber.}}
\label{fig:cost-sbs-ratio}
\end{figure}

Fig.~\ref{fig:cost-sbs-ratio} presents the cost as the number of SBSs connected with fiber. This plot compares six distinct strategies: FBCP focusing on the deployment cost ($\alpha$=0.0), random (RND), and four spatial diversity-based FBCP schemes defined by $\alpha$ parameters ranging from 0.25 to 1.0. All strategies start at a base cost of approximately 5,600 units when no SBS is fiber-connected.

The FBCP($\alpha$=0) strategy, which prioritizes the least expensive SBSs to connect, consistently yields the lowest cost across all connection levels. At 40 connected SBSs, the cost incurred is about 13,300 units under FBCP($\alpha$=0), compared to approximately 19,600 units for the RND strategy, and over 20,750 units for the most spatially diverse configuration FBCP($\alpha$=1). This demonstrates that enforcing spatial diversity in SBS placement while potentially beneficial for coverage can increase deployment cost by more than 60\%.
All strategies converge to a total cost of around 24,000 as they are using the same long-term fiber placement planning strategy according to Algorithm~\ref{alg:connected-topology}.

% Interestingly, a crossover point is observed near 70 connected SBSs, where both the FBCP($\alpha$=0) and RND strategies converge to a total cost of around 24,000 units, which also corresponds to the upper bound when all 79 SBSs are fiber-connected. This convergence reflects the diminishing impact of selection strategy as the network approaches full deployment.

\subsection{Hybrid Fiber-FSO Backhaul Strategies}

\begin{figure}[t]
\centering
\includegraphics[width=\linewidth, clip=false]{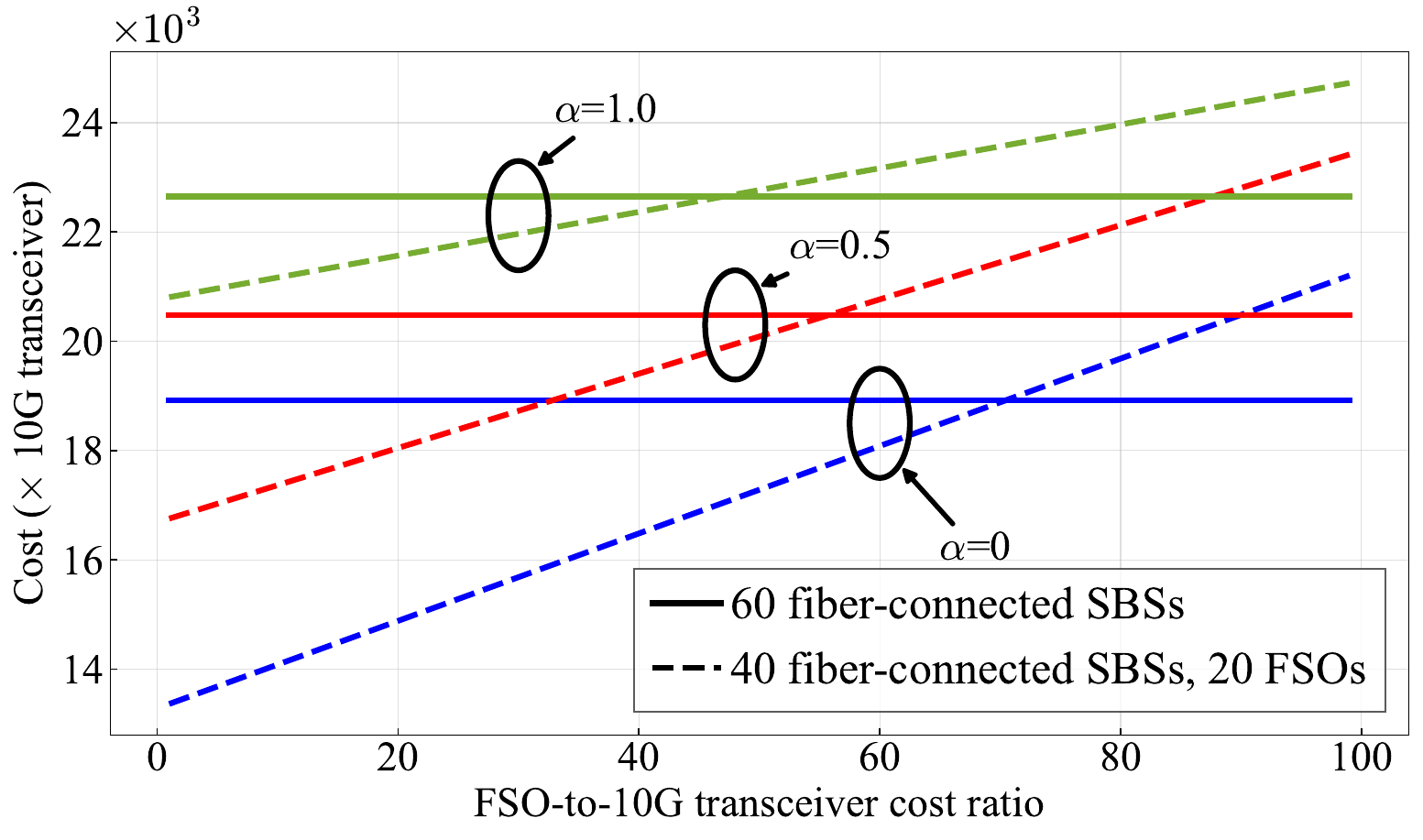}
\caption{\textcolor{black}{Cost as a function of the FSO-to-10G transceiver cost ratio for 60 connected SBSs.}}
\label{fig:cost-fso-ratio}
\end{figure}

Fig.~\ref{fig:cost-fso-ratio} illustrates the total cost of connecting 60 SBSs as a function of the FSO-to-10G transceiver cost ratio $\beta_{FSO}$. 
Two scenarios are examined: (i) all 60 SBSs are fiber-backhauled (solid lines), and (ii) 40 SBSs are fiber-backhauled while 20 SBSs utilize FSO backhaul (dashed lines). These scenarios are evaluated for three FBCP methods with $\alpha$ values of 0.0, 0.5, and 1.0. 
For each FBCP method, a crossover point is evident, indicating the $\beta_{FSO}$ value beyond which connecting additional SBSs with FSO incurs a lower cost than using fiber. 
For FBCP($\alpha$=0) , which solely prioritizes cost, this crossover occurs at approximately $\beta_{FSO}$=70. 
n contrast, for the  FBCP($\alpha$=1), method, which emphasizes spatial separation, the crossover point appears at a lower $\beta_{FSO}$ value of around 47, suggesting that a focus on cost minimization also indirectly mitigates the higher relative cost of FSO backhauling.

\begin{figure}[t]
    \centering
    \begin{minipage}{1\linewidth}
    % \subcaption{\raggedright}
    % \vspace{-.5cm}
    \hspace{.5cm}\includegraphics[width=.95\linewidth]{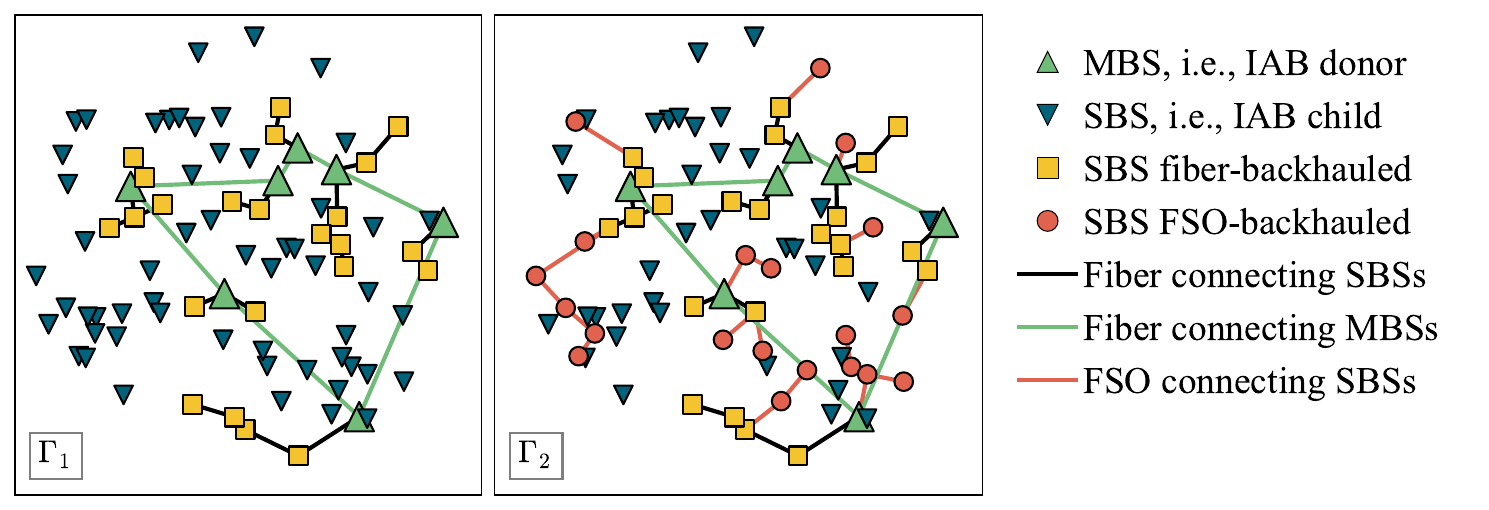}
    \end{minipage}
    \begin{minipage}{1\linewidth}
    \subcaption{\raggedright}
    \vspace{-.5cm}
    \hspace{.5cm}\includegraphics[width=.95\linewidth]{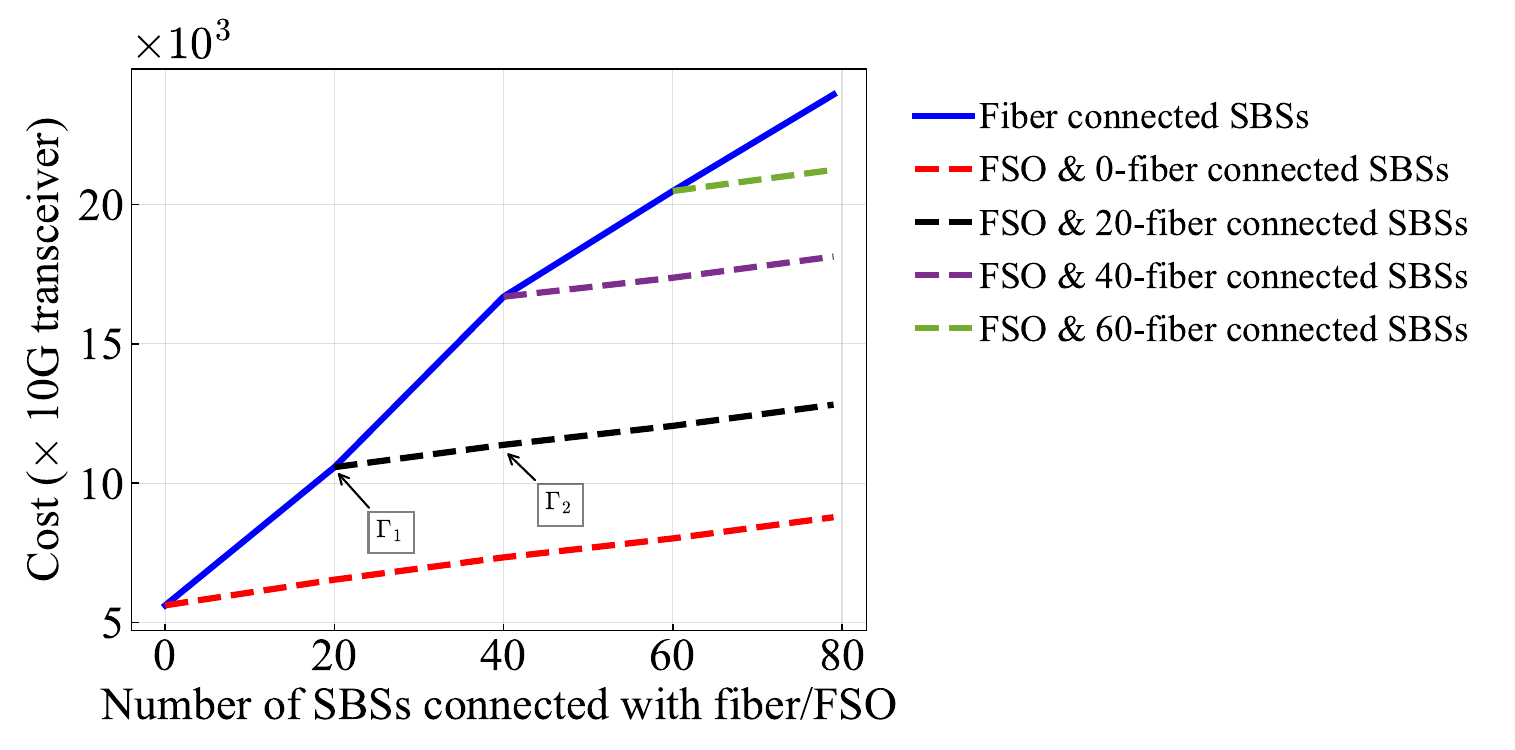}
    \end{minipage}
    \caption{\textcolor{black}{Cost and corresponding topologies for SBSs connected with fiber and FSOs for FBCP($\alpha$=0.5); (a) Topologies connecting 20 SBSs with fiber (left); 20 SBSs with fiber and 20 SBSs with FSOs (right); (b) Cost as a function of the number of SBSs connected with fiber and FSOs}}
    \label{fig:cost-fso-topologies}
\end{figure}

Fig.~\ref{fig:cost-fso-topologies} (top) demonstrates the topologies connecting fiber and FSOs in the backhaul, while Fig.~\ref{fig:cost-fso-topologies} (bottom) examines the cost efficiency of hybrid deployments for FBCP($\alpha$=0.5), as a function of connected SBSs. The blue solid line (\textit{Fiber connected}) represents the scenario where all connected SBSs are fiber-backhauled.
The other dashed lines depict cases with a fixed number of fiber-connected SBSs, with the remaining SBSs being FSO-connected.
The fully fibered scenario reaches the maximum cost of approximately 24,00 units, while the fully FSO-based configuration (\textit{FSO \& 0-fiber connected}) has the lowest cost, roughly 8,800 units. 
Hybrid approaches show nearly linear cost growth. When 20 SBSs are fiber-connected (and remaining 59 are FSOs connected), the total cost rises to around 12,800 units. This increases to 18,100 units for 40 fiber-connected SBSs, and to approximately 21,200 units at 60. 
The top of the figure illustrates the placement of fiber and FSO links for two cases: (i) 20 SBSs connected solely with fiber, and (ii) 20 SBSs fiber-backhauled and 20 SBSs FSO-backhauled. 
% The curve for 79 fiber-connected SBSs effectively a near-complete fiber rollout approaches 23,700 units, nearly matching the full fiber cost.
These trends highlight the cost benefit of selective fiber deployment. For example, opting for a hybrid model with 40 fiber-connected SBSs can yield over 30\% cost savings compared to a fully fibered network, while enabling strategic use of reliable links where most needed.

% \begin{figure}[h!]

% \begin{subfigure}{\textwidth}
% \captionsetup{singlelinecheck = false, format= hang, justification=raggedright, labelsep=space} 
%     \includegraphics[width=3.5in\textwidth]{Figures/illustration_fig1.png}
%     \caption{subplot a.}
%     \label{fig:second}
% \end{subfigure}
% \begin{subfigure}{\textwidth}
% \captionsetup{singlelinecheck = false, format= hang, justification=raggedright, labelsep=space}    
%     \includegraphics[width=3.5in\textwidth]{Figures/Plot1_Line.png}
    
%     \caption{subplot b.}
%     \label{fig:third}
% \end{subfigure}
        
% \caption{Service coverage probability as a function of the distance from IAB donor to child IAB $s$ in subplot a with blockage $\lambda_{\text{bl}}$ = 500 $\text{km}^{-2}$.}
% \label{linebs1}
% \end{figure}

% \begin{figure}
% \centerline{\includegraphics[width=3.5in]{Figures/fiber_to_fso_cost.pdf}}
% \caption{\textcolor{black}{Cost as a function of the number of SBSs connected with fiber and FSOs}}
% \label{fso}
% \end{figure}

\begin{figure}
\centerline{\includegraphics[width=\linewidth]{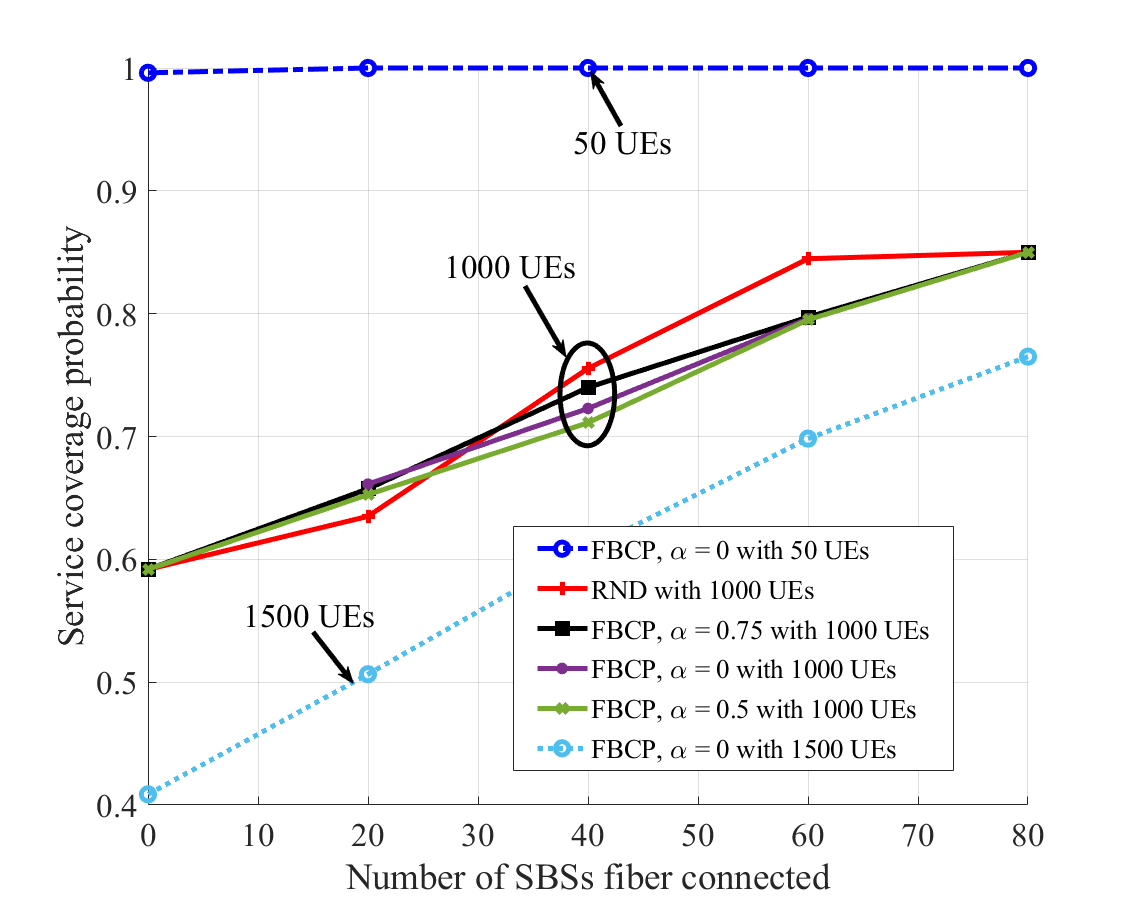}}
\caption{Service coverage probability as a function of the number of SBSs connected with fiber, $P_{\text{MBS}}$ = \text{40 dBm}, $P_{\text{SBS}}$ = \text{24 dBm}}
\label{service_coverage_sbs}
\end{figure}

\begin{figure}
\centerline{\includegraphics[width=\linewidth]{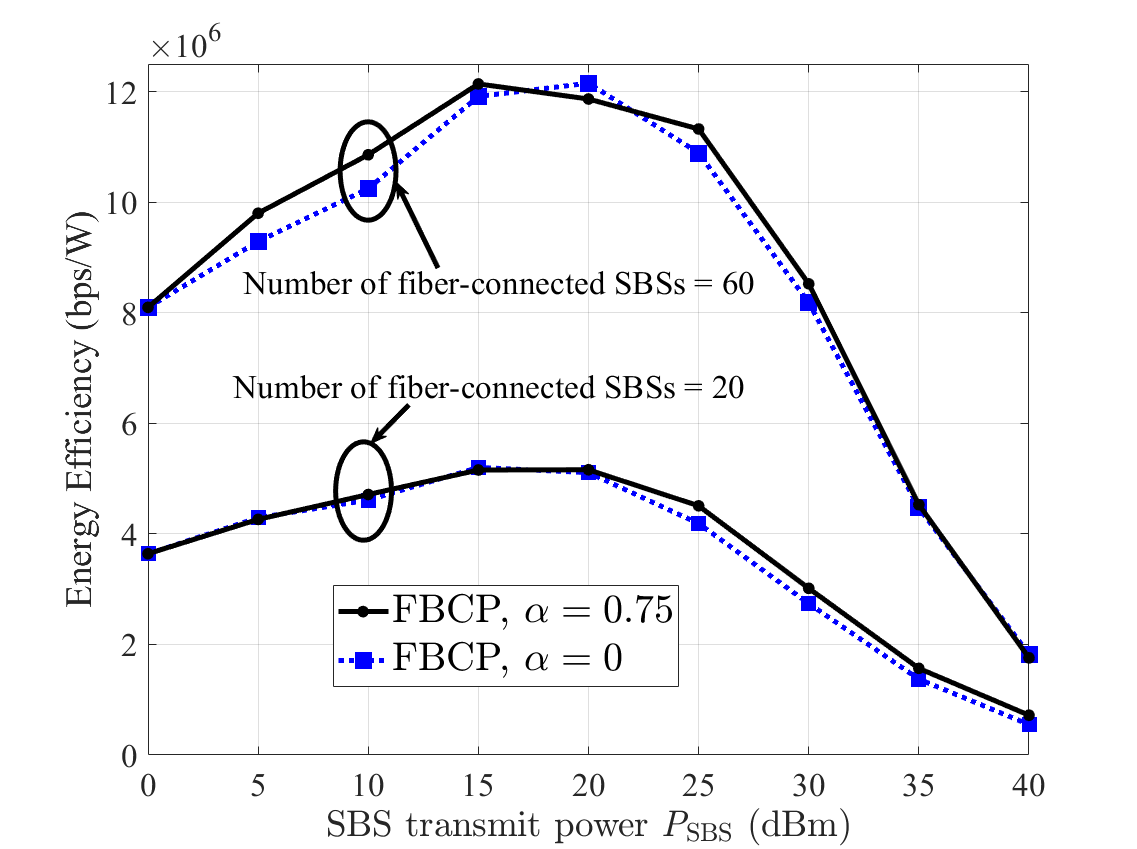}}
\caption{Energy efficiency as a function of the
SBS transmit power with $P_{\text{MBS}}$ = \text{40 dBm}, power loss in fiber at 1W/km, and the power consumption of a fiber transceiver at 8W.  }
\label{energy_efficiency}
\end{figure}

\subsection{Impact of Fiber-Connected SBSs on Service Coverage Probability}

In Fig. \ref{service_coverage_sbs}, we present the impact of the number of fiber-connected SBSs on service coverage probability under various deployment strategies. As we see, the \textit{cost sort}, i.e., $\alpha$=0) scheme for low UE density
(50 UEs) achieves nearly ideal performance, maintaining
a service coverage probability of 0.99 or higher regardless of the number of fiber-connected SBSs.

As the UE density increases from 1,000 to 1,500, the service coverage probability decreases, however, connecting more SBSs with fiber gradually increases performance.
For instance, with 1,000 UEs, the coverage improves from approximately 0.4 with no fiber-connected SBSs to over 0.76 with 80 SBSs connected. These trends show the scalability and efficiency of the cost-aware SBS selection approach.

In contrast, random SBS selection with 1000 UEs performs sub-optimally, particularly when smaller number of SBSs are fiber-connected out of the possible 80. However, when the number of SBSs fiber-connected surpasses about 26-30, the service coverage probability outperforms other deployment strategies. 
When the number of SBSs to be fiber-connected is small (e.g., 26-30), the number of possible random configurations is high. This high variability leads to a wider range of possible outcomes, including many suboptimal configurations where, fiber-connected SBSs may be too concentrated in certain areas and edges, leaving coverage gaps. Some SBSs with high UE density may not be selected for fiber connection, resulting in poor service coverage. However, when the number of fiber-connected SBSs increases, it results in a reduced configuration space. Since fewer SBSs are remaining to choose from, the random selection becomes less variable, and the likelihood of forming effective configurations increases. Also, with more SBSs connected to fiber, the network is closer to achieving coverage saturation, meaning that most critical areas already have a fiber-connected SBS. The spatial separation strategies ($\alpha = 0.5$ and $\alpha = 0.75$) align closely with that of \textit{cost sort}, i.e., $\alpha = 0$. While increasing $\alpha$ introduces spatial diversity, it often incurs higher deployment cost, making the deployments less favorable compared to \textit{cost sort}, i.e., $\alpha = 0$ scenario. As we see, targeted FBCP deployment, potentially augmented by spatial considerations, emerges as a robust strategy for maintaining service coverage probability.

\subsection{Impact on the energy efficiency}

Fig.~\ref{energy_efficiency} depicts the network energy efficiency, measured in bits per second per watt (bps/W), as a function of the transmit power of SBSs, \(P_{\text{SBS}}\), across four different deployment scenarios. These include \textit{cost sort}, i.e., $\alpha$ = 0, strategy with 20 and 60 SBSs that are fiber-connected, and FBCP(\(\alpha=0.75\)), using the same SBS counts.

Here, the energy efficiency is given by $ \text{EE} = \frac{R_{\text{total}}}{P_{\text{total}}} $, where the total achievable system throughput is represented by $R_{\text{total}}$, while the total power consumed across the network is denoted by $P_{\text{total}}$, which can be expressed as:
\begin{equation}
P_{\text{total}} = P_{\text{MBS}} + P_{\text{SBS, wireless}} + P_{\text{SBS, fibered}}.
\end{equation}
In all configurations, energy efficiency increases initially with \(P_{\text{SBS}}\), peaks around 15-20 dBm, and then decreases. This trend highlights the trade-off between increased throughput and rising power consumption. As transmit power increases, user data rates improve due to stronger desired signals, but the associated increase in network power consumption eventually dominates, reducing overall efficiency.

Among the schemes, separation-based deployment (\(\alpha=0.75\)) with 60 SBSs achieve the highest energy efficiency, peaking above \(1.2 \times 10^7\) bps/W. The \textit{cost sort} strategy (when $\alpha$ = 0) with 60 SBSs also performs comparably well, indicating that optimized SBS placement either by cost or spatial diversity yields energy-efficient performance. In contrast, the $\alpha$ = 0 and \(\alpha=0.75\) deployments with only 20 SBSs demonstrate lower efficiency. This highlights the role of SBS density in achieving energy-efficient coverage, as fewer nodes lead to higher per-node loads and less spatial reuse. Overall, we see that moderate transmit powers combined with intelligent SBS selection strategies (cost-aware or spatially diverse) can significantly boost energy efficiency in dense small-cell networks, of IAB type.

\section{Conclusion}

We studied the problem of IAB network infrastructure optimization with fiber/FSO assisted backhaul. 
%in the cases with different fiber layout strategies and hybrid FSO deployments. 
We proposed a fiber backhaul connection placement algorithm for planning fiber rollout according to separation vs. cost weighting parameter.
%...... methods with no need for mathematical analysis and with the capability to be adapted for different channel models/constraints/metrics of interest. 
As demonstrated, proper backhaul network planning can  boost the IAB networks' coverage, reduce the cost of deployment, and energy efficiency. In practice, factors such as the availability of non-IAB backhaul connections in certain regions and local regulatory constraints may significantly impact the design process. Additionally, designers may need to account for planned infrastructure modifications, regional cost considerations, and seasonal fluctuations.

\section*{Acknowledgment}

This work was supported by the ECO-eNET project with funding from the Smart Networks and Services Joint Undertaking (SNS JU) under grant agreement No. 10113933. The JU receives support from the European Union's Horizon Europe research and innovation.

\bibliographystyle{IEEEtran}
\bibliography{bibliography}

\end{document}